\documentclass[review, pdftex]{elsarticle}

\usepackage{hyperref}

\journal{Journal of \LaTeX\ Templates}

\usepackage{amsmath}
\newcommand{\bhline}[1]{\noalign{\hrule height #1}}

\begin{document}

\begin{frontmatter}

\title{Development of an agent-based speculation game for higher reproducibility of financial stylized facts}


\corref{mycorrespondingauthor}\author[mymainaddress,mysubaddress]{Kei Katahira}
\cortext[mycorrespondingauthor]{Corresponding author.}
\ead{k.katahira@scslab.k.u-tokyo.ac.jp}

\author[mymainaddress]{Yu Chen}
\author[mymainaddress]{Gaku Hashimoto}
\author[mymainaddress]{Hiroshi Okuda}

\address[mymainaddress]{Graduate School of Frontier Sciences, The University of Tokyo, 5-1-5 Kashiwanoha, Kashiwa-shi, Chiba-ken 277-8563, Japan}
\address[mysubaddress]{Research Fellow of Japan Society for the Promotion of Science}

\begin{abstract}
Simultaneous reproduction of all financial stylized facts is so difficult that most existing stochastic process-based and agent-based models are unable to achieve the goal. In this study, by extending the decision-making structure of Minority Game, we propose a novel agent-based model called ``Speculation Game,'' for a better reproducibility of the stylized facts. The new model has three distinct characteristics comparing with preceding agent-based adaptive models for the financial market: the enabling of nonuniform holding and idling periods, the inclusion of magnitude information of price change in history, and the implementation of a cognitive world for the evaluation of investment strategies with capital gains and losses. With these features, Speculation Game succeeds in reproducing 10 out of the currently well studied 11 stylized facts under a single parameter setting. 
\end{abstract}

\begin{keyword}
Econophysics, Multi-agent simulation, Financial stylized facts, Cognitive model, Round-trip trading 
\end{keyword}

\end{frontmatter}


\section{Introduction}
\subsection{Background}
{\it Stylized facts} are qualitative properties of asset returns, which are indicated as the result of more than half a century of empirical statistical studies on financial time series \cite{cont2001empirical}. The 11 currently reported stylized facts can be listed as follows: volatility clustering, intermittency, heavy tails, the absence of autocorrelation in returns, slow decay of autocorrelation in volatilities, volume/volatility correlation, aggregational Gaussianity, conditional heavy tails, asymmetry in time scales, leverage effect and gain/loss asymmetry. They are quite nontrivial features which have been well studied for many years in different markets and for various instruments. 

The reproducibility of the stylized facts is a prerequisite for the financial market model. However, to reproduce them with a conventional market model, whether a simple or sophisticated stochastic process, is not very easy due to the high level of multiplexity of the market. With the contribution of econometrists, several well known stochastic models have been constructed to yield typical stylized features, for example, the autoregressive conditional heteroscedasticity (ARCH) and generalized autoregressive conditional heteroscedasticity (GARCH) processes \cite{bollerslev1986generalized, mantegna1999introduction}. Nonetheless, to the best of our knowledge, most existing models fail to reproduce the whole set of stylized properties simultaneously.  

On the other hand, an alternative approach mainly advocated by econophysicists, which adopts agent simulation based on a set of simple rules, has been used to study the stylized facts as well. The simulation enables the reproduction and analysis of complex price dynamics in an artificial market with autonomous agents acting as the traders. Compared with previous stochastic process models, which are designed to describe the traits of financial time series at the macroscopic level, this method is a bottom-up approach which explains these features at the microscopic level, that is, the level at which traders' decision-making and trading behaviors are taking into account. 

There are mainly two reasons for adopting agent-based models. First, the agent-based simulation is almost the only way to analyze the effects of traders' behavior to the whole market. All time-series data obtained from the real market, such as price and volume, are aggregational ones, which is generally impossible to be used in tracking the individual actions of all market participants. Stochastic process models have a high ability to address macroscopic phenomena, but analyses based on such models ignore those endogenous factors, such as mutual interaction among traders. Second, multi-agent simulation can offer a scenario that explains the generation process of emergent phenomena, which were essentially regarded as a black box in the traditional financial analyses. On the other hand, analyses with agent simulation can provide, at least partially, a reasonable scenario for such a complex problem. 

As physicists tend to believe that all phenomena in this world could be subject to some simple universal rules at the basic level, several ``simple'' agent-based models have been proposed. Specifically, these simple models have the following characteristics:
\begin{enumerate}
\item A bottom-up structure.  
\item Elements (agents) governed by simple rules appropriately abstracted from real traders' behavior.
\item Aggregational output (e.g., price change) in statistical agreement with the real data.
\end{enumerate}
In particular, the second point is important because the simplicity of agents' rules allows relatively easier analysis of mechanisms responsible for the complicated price dynamics. A tractable model like such is called a {\it toy model}.

Note that the agent-based model proposed by economists often loosens the requirement of simplicity in the microstructure and pursues agents' behavioral rules as sophisticated as possible. Examples of such a complex agent-based market model are the Santa Fe Institute model \cite{palmer1994artificial, arthur1996asset, lebaron1999time} and the Genoa artificial stock market model \cite{raberto2001agent, raberto2003traders}. Compared with toy models, these complex agent-based market models have higher reproducibility of stylized facts, but they are more difficult to use in identifying critical factors at the microscopic level and in linking these factors to the aggregational outputs. 

Typically, toy models for financial markets can be divided into two groups depending on whether a learning mechanism is included: zero-intelligence models and adaptive models. The representative zero-intelligence models are the percolation model \cite{cont2000herd, stauffer1998crossover}, Ising (spin) model \footnote{Some Ising models such as \cite{sornette2006importance, zhou2007self} also incorporate learning mechanisms so that they could be classified as the adaptive model as well.} \cite{sornette2006importance, zhou2007self, kaizoji2002dynamics, sornette2014physics}, empirical order-driven model \cite{mike2008empirical, gu2009emergence, meng2012effects, zhou2017computational}, and Sznajd model \footnote{It was originally a mathematical model for opinion formation. The use of this model for financial markets is one example of its applications \cite{sznajd2005sznajd}.} \cite{sznajd2002simple}. The famous adaptive models include the Lux-Marchesi model \cite{lux1999scaling}, Minority Game \cite{challet1997emergence, challet2005minority} and its extended versions, such as the grand canonical Minority Game (GCMG) \cite{challet2001stylized, challet2001games, challet2003criticality}, \$-game \cite{andersen2003game}, and pattern game \cite{challet2008inter}. Previous studies using these agent-based toy models achieved remarkable progress in the analysis of market dynamics. For example, the family of Minority Game typed models revealed that the stylized facts of price returns can arise as long as agents have possibility to adaptively modulate their investment. Particularly, this is understood as a combination of inductive learning with dynamic capitals \cite{challet2001minority, galla2009minority} or a switch between `active' and `inactive' strategies depending on their performance \cite{challet2001stylized, bouchaud2001universal}, both of which would result in a {\it herding behavior} among traders \cite{manuca2000structure}. 

Meanwhile, toy models might also overlook some fundamental factors and fail in reproducing the related stylized facts. For instance, the payoff strategy of Minority Game results in a mean-reverting aggregated outcome, which is a main cause for the loss of price diffusivity. This feature might be relieved if the investing strategies were evaluated with capital gains and losses by round-trip trades, as Katahira and Akiyama pointed out in their study \cite{katahira2017}. In our opinion, a toy model that allows a little more sophistication, but not to the extent of the complex agent-based market models, has the potential to enhance the reproducibility of stylized facts, and at the same time retain the advantage of being able to find the latent mechanisms easily. 

\subsection{Purpose}
In this research, we build an agent-based toy model for the financial markets, named ``Speculation Game,'' with more realistic factors in order to obtain a better reproducibility of the stylized facts. In particular, three extensions are made to preceding models: the enabling of nonuniform {\bf holding and idling periods}, the inclusion of {\bf magnitude information of price change} to history, and the implementation of a {\bf cognitive world} for the evaluation of investment strategies with capital gains and losses. The focus of our study is not merely the perfect reproduction of all the stylized facts, but the fulfillment of the prerequisite as a useful financial market model for Speculation Game. In fact, elucidation of the emerging mechanism of stylized facts is more crucial than the reproduction itself, which will be the main theme of our next paper. 

The remainder of this paper is organized as follows. The next section describes the model, and Section 3 presents the results and discussion. Finally, Section 4 summarizes the study.

\section{Model}
\subsection{Model design}
As the kind of {\it bounded rationality} discussed by W. Brian as ``The El Farol Bar Problem'' \cite{arthur1994inductive} is implemented, Minority Game can be considered as one of the most insightful toy models regarding human inductive behavior for the complex adaptive system including financial markets. An abstract decision-making process is provided for each player in the game with the specifications of {\bf history}, {\bf memory}, and {\bf strategy}. 

Speculation Game is constructed by exploring the decision-making structure of Minority Game. To develop our model, these structures are utilized and a minimum set of rules enabling the round-trip trading is designed. 

\subsection{Gaming system}
In Speculation Game, $N$ players participate a gamified market, competing with each other to increase their wealth by capital gains through round-trip trades \footnote{Ferreira and Marsili \cite{ferreira2005real} attempt to introduce a round-trip trade into Minority Game, but it is restricted as a successive two-step transaction. The pattern game proposed by Challet \cite{challet2008inter} incorporates round-trip payoffs by taking account of the reaction time of order placement.}.  

At each time step $t$, player $i$ takes an action $a_i(t)$ from three alternatives \footnote{Chow and Chau \cite{chow2003multiple} analyze Minority Game in which every player has more than two choices.}---buy ($=1$), sell ($=-1$), and hold (idle) ($=0$)---based on information of past price movements, called history \footnote{History in Minority Game records the past minority groups.}. Note that action $0$ is explained as ``hold''  during an ongoing round-trip trade, and as ``idle'' during the waiting period before the opening of a new position. The history of price change is represented as a quinary time series $H(t)$: largely up ($=2$), up ($=1$),  stay($=0$), down ($=-1$), and largely down ($=-2$). The player with memory $M$ can refer to the last $M$ digits of $H(t)$, shown as Equation \ref{eq1}, in which $h(t)$ is a quantized price movement at time step $t$ as, 
\begin{equation}
H(t) = (h(t-M),\cdots,h(t-1)).
\label{eq1}
\end{equation}
Note that $H(0)$ is randomly initialized.

When a trading decision is to be made, every player uses a strategy table to determine recommended action corresponding to the historical patterns (see Table \ref{tab1}). The total number of strategies is $3^{5^M}$ since players have three choices and there are $5^M$ possible histories. For the realization of round-trip trades, a player will take ``hold'' action either if action $0$ is recommended by the strategy table, or if the recommendation ($1$ or $-1$) is the same as the action in opening positions. This rule enables each player to manage the timing of opening and closing positions in reference with buying and selling signals encoded in price movements. Under such a constraint for the completion of round-trip trades, players' actions are not only determined by investing strategies but also by their trading statuses (i.e., opened or closed positions). This situation will impede the synchronized actions among the players (see Appendix A for details). 
\begin{table}[htbp]
\caption{Example of a strategy for $M=3$.}
\label{tab1}
\begin{center}
\begin{tabular}{rrr|rrrrr} \bhline{1.1pt}
\multicolumn{3}{c|}{History} &  \multicolumn{5}{c}{Recommended action}\\ \bhline{1.1pt}
$-2$ & $-2$ & $-2$ & & & & $1$ &\\
$-2$ & $-2$ & $-1$ & & & & $0$ &\\
$-2$ & $-2$ & $0$ & & & & $0$ &\\
$-2$ & $-2$ & $1$ & & & & $-1$ &\\
$-2$ & $-2$ & $2$ & & & &  $1$ &\\
$-2$ & $-1$ & $-2$ & & & &  $0$ &\\
&  \vdots \,  &  & & & & \vdots \, & \\
$2$ & $2$ & $2$ & & & & $-1$ & \\ \bhline{1.1pt}
\end{tabular}  
\end{center}
\end{table}

In addition, while a genuine or a fake history is not essential to proceed Minority Game \cite{cavagna1999irrelevance}, a history generated from the actual price changes is required in Speculation Game for the players to find out profitable opportunities, which will largely influence the aggregational results. With these significant differences from Minority Game, Speculation Game can be regarded as a different type of agent-based model although the basic elements such as history, memory, and strategy tables are borrowed from it.

Player $i$ of Speculation Game possesses randomly chosen $S$ strategies and uses his or her most successful strategy in terms of the accumulated strategy gains $G_i^j(t)$ ($j \in S$) at each time step. In addition, unused strategies are evaluated as if they were used. Whenever a trade is closed by the strategy in use, performances of all the strategies are reviewed to find the one with the best performance. If the best strategy happens to be the current one, the player continues to use it. However, if the best strategy turns out to be a strategy that is not in use (i.e., different from the current strategy) with which a virtual trade might be still ongoing, the virtual position is cleared immediately and the player switches to use it at the next time step. Note that when a virtual round-trip trade is aborted by switching the best strategy, the accumulated strategy gains $G_i^j(t)$ will not be updated.

Players in Speculation Game can place an order with variable trading volumes \footnote{GCMG introducing evolving capital with variable investments are studied in \cite{challet2001minority, galla2009minority, jefferies2001market, johnson2003financial}. The Patzelt-Pawelzik model \cite{patzelt2013inherent} also takes a similar approach.}. The quantities of order which is allowed for player $i$ to place depends on his or her market wealth $w_i(t)$ as follows,
\begin{equation}
q_i(t) = \lfloor \frac{w_i(t)}{B} \rfloor.
\label{eq2}
\end{equation}
$B$ is the {\bf board lot amount}, which describes the ease of placing orders with multiple quantities and is a significant parameter for the dynamical evolution of the players' market wealth. Equation \ref{eq2} implies that a player with ample wealth (e.g., a big speculator, such as a hedge fund in the real market) can place an order with massive quantities. Note that opening and closing volumes of a round-trip trade are required to be the same. Moreover, the amount of initial market wealth for each player is decided by a uniformly distributed random number $U[0,100)$ as 
\begin{equation}
w_i(0) = \lfloor B+U[0,100) \rfloor,
\label{eq3}
\end{equation}
in order to enable the player to place a unit volume order at least.

For market price change $\Delta p$, which will be used in formulating the decision making process of players, we simply follow the order imbalance equation defined as the function of excess demand $D(t)$ in the reference of Cont and Bouchaud \cite{cont2000herd}, 
\begin{equation}
\Delta p = p(t) - p(t-1) = \frac{D(t)}{N},
\label{eq4}
\end{equation}
where $N$ works as the ``depth'' of game market. While the log-return of price is adopted in such an order imbalance relationship for GCMG models \cite{challet2001stylized, challet2001games, challet2003criticality}, we assume that the price change is directly driven by the excess demand because a cognitive threshold is introduced in Speculation Game as shown in Equation \ref{eq6} \footnote{There is no absolute rule with which the price change or the log-return of price should be applied to this kind of order imbalance equation. Indeed, the price change is used in the Cont-Bouchaud percolation model \cite{stauffer1998crossover}.}. The reason is that, for an ordinary trader in the real market, it is not natural to set a cognitive threshold for log-return, which would require an extra calculation from the price chart, instead of price change. In the test simulations, if log-return were used for the price formations, some extreme fluctuations of market price could be observed due to the loss of function for the cognitive threshold.

The initial market price can be arbitrarily chosen to a certain extent as long as it is large enough to avoid the appearance of negative prices. In this paper, it is set as $p(0)=100$. Also, the excess demand $D(t)$ is calculated with the quantities $q_i(t)$ and also the actions $a_i^{j^*}(t)$ recommended by strategy $j^*$ in use:
\begin{equation}
D(t) = \sum_{i=1}^{N}a_i^{j^*}(t)q_i(t). 
\label{eq5}
\end{equation}
Furthermore, the definition of price change in Equation \ref{eq4} implies the presence of a market maker who provides sufficient liquidity in the game market; hence, the excess demand can always be balanced instantly by a counterpart. 

\begin{figure}[htbp]
\begin{center}
\includegraphics[width=1.0\textwidth]{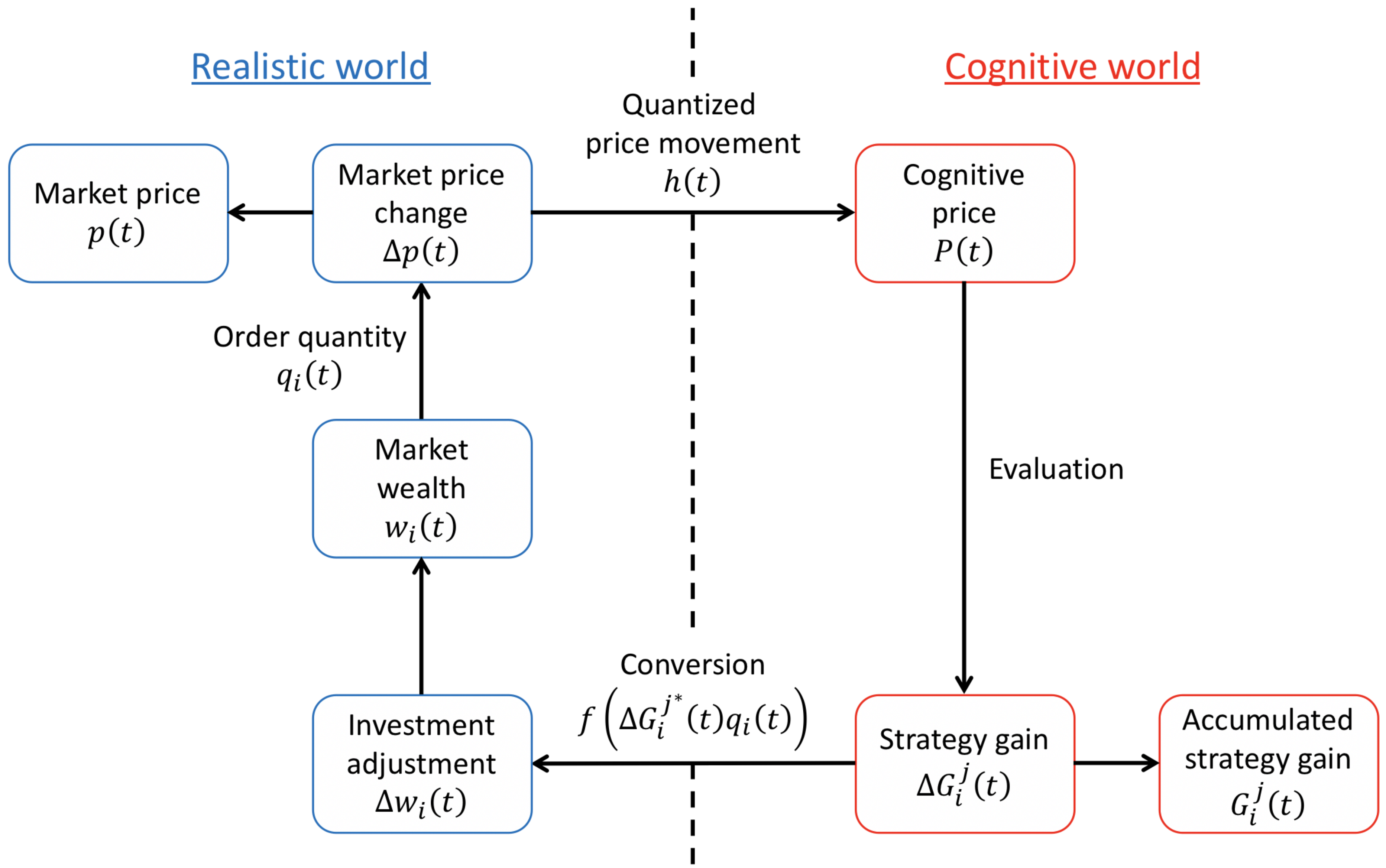}
\end{center}
\caption{The mutual projection between the realistic and cognitive worlds in Speculation Game.}
\label{fig1}
\end{figure}

The quantized price movement $h(t)$ is determined based on the definition of $\Delta p$ in Equation \ref{eq4} as follows:
\begin{equation}
h(t) = \begin{cases}
2 & (\Delta p>C)\\
1 & (C \geq \Delta p>0)\\
0 & (\Delta p=0)\\
-1 & (-C \leq \Delta p<0)\\
-2 & (\Delta p<-C)
\end{cases}.
\label{eq6}
\end{equation}
The {\bf cognitive threshold} $C$ is used to discriminate the significant price changes (recognized as a big move by the players) from insignificant ones. Since the cognized history $H(t)$ only contains very rough information about the magnitude of $\Delta p$, a kind of designated coarse price $P(t)$, instead of the market price $p(t)$, needs to be prepared for the players to make their decisions properly. Note that Speculation Game would not work well if players evaluated their strategies directly with the market price. Indeed, no player could make a profit in that case because of the inconsistency between the preciseness of information in the cognized history and that in the market price. Hence, an assumption is made that players calculate their gain or loss of investment only in their cognitive worlds, where also their trading decisions are made. This assumption (i.e., the evaluation with simplified payoffs rather than real ones) is consistent with the empirical study done by De et al. \cite{de2010success}, which reveals that investors have a cognitive bias such that their current trading decisions are more influenced by the sign of outcomes of past trades (positive or negative) rather than the actual size of those. In the current study, $P(t)$ is defined as the cognitive price, which is calculated as follows, letting $P(0) = 0$,
\begin{equation}
P(t) = P(t-1)+h(t).
\label{eq7}
\end{equation}
Accordingly, the players' strategy gains are calculated based on this cognitive price. If player $i$ with strategy $j$ opens a position at time step $t_0$ and closes it at time step $t$, his or her strategy gain in this round-trip trade is: 
\begin{equation}
\Delta G_i^j(t) = a_i^j(t_0)(P(t) - P(t_0)).
\label{eq8}
\end{equation}
The performance of strategy $j$ of player $i$ is measured by its accumulated strategy gain as follows:
\begin{equation}
G_i^j(t) = G_i^j(t_0) + \Delta G_i^j(t).
\label{eq9}
\end{equation}
The player's market wealth $w_i(t)$ is updated in a similar way with the order quantities $q_i(t)$ taken into consideration. When the strategy in use is $j^*$, $\Delta G_i^{j^*}(t)$ shall be converted into the player's investment adjustment as 
\begin{equation}
\Delta  w_i(t) = f(\Delta G_i^{j^*}(t)q_i(t_0)), 
\label{eq10}
\end{equation}
where $f$ can be an arbitrary monotonically increasing function. In the following texts, $\Delta  w_i(t) = \Delta G_i^{j^*}(t)q_i(t_0)$ is used for simplicity. Note that this $\Delta w_i(t)$ would be different from the one calculated directly based on the market price. Here, we assume the difference could be adjusted externally. In other words, self-finance is not postulated in Speculation Game. After the adjustment, the market wealth of player $i$ is updated as follows:
\begin{equation}
w_i(t) = w_i(t_0) + \Delta  w_i(t).
\label{eq11}
\end{equation}
As the difference between Equations \ref{eq9} and \ref{eq11} shows, the unit volume gain is applied for the calculation of strategy performance. This is because strategies ought to be evaluated only in terms of the timing of order placement. The framework of the model explained so far is summarized in Figure \ref{fig1}. As the figure shows, Speculation Game proceeds with the players behaving alternately in the realistic and in the cognitive worlds.

In addition, each player enters or exits the game market depending on his or her market wealth. When $w_i(t)<B$, which means that the player does not have enough wealth to order even a minimal unit, then the player has to withdraw from the market. In this case, an alternative player will enter as a substitute with a new set of $S$ strategies randomly given, and an initial market wealth generated by $\lfloor B+U[0,100) \rfloor$.  

To sum up, in a different way from Minority Game typed models, both the concepts of `active' and `inactive' of players \cite{challet2001stylized, bouchaud2001universal} and dynamic capitals \cite{challet2001minority, galla2009minority} are implemented in this model by realizing round-trip trades with variable order quantities.

\section{Results and discussions: Reproducibility of stylized facts}
In order to assess Speculation Game for the reproducibility of the stylized facts, we calculate the market price return as $r(t) = {\rm ln}(p(t)) - {\rm ln}(p(t-1))$ (to follow the analysis method in previous studies) and study the 11 specific stylized facts reported by Cont \cite{cont2001empirical}, which have been acknowledged for a long time as the general qualitative characteristics of financial markets. The simulations are conducted under a fixed parameter setting: $N=1000$, $M=5$, $S=2$, $B=9$, $C=3$. This baseline calculation only serves as a demonstration of Speculation Game; stylized facts can also be reproduced under other parameter settings, which will be detailed in another paper. 

\subsection{Volatility clustering and intermittency}
Volatility clustering is the tendency of large changes in price returns to cluster together \cite{cont2001empirical, challet2001stylized, rydberg2000realistic}. In other words, price fluctuates violently once volatility, generally measured by $|r(t)|$ or $r(t)^2$, increases and vice versa. This feature introduces heavy tails in the distribution of returns and has a strong relationship to the slow decay of autocorrelation in volatilities.

Meanwhile, intermittency describes the presence of intermittent bursts at any time scale in the time series of price returns. In other words, a phase with large volatilities alters chaotically with a phase with small volatilities, with varying lengths of periods of these interchanges. This kind of aperiodic instability can also be observed in several physical systems and known as on-off intermittency in the physics literature \cite{lux1999scaling, platt1993off}. Mantegna and Stanley \cite{mantegna1999introduction} indeed noticed the similarity between the price changes in financial markets and the velocity fluctuations in turbulence.

\begin{figure}[htbp]
\begin{center}
\includegraphics[width=1.0\textwidth]{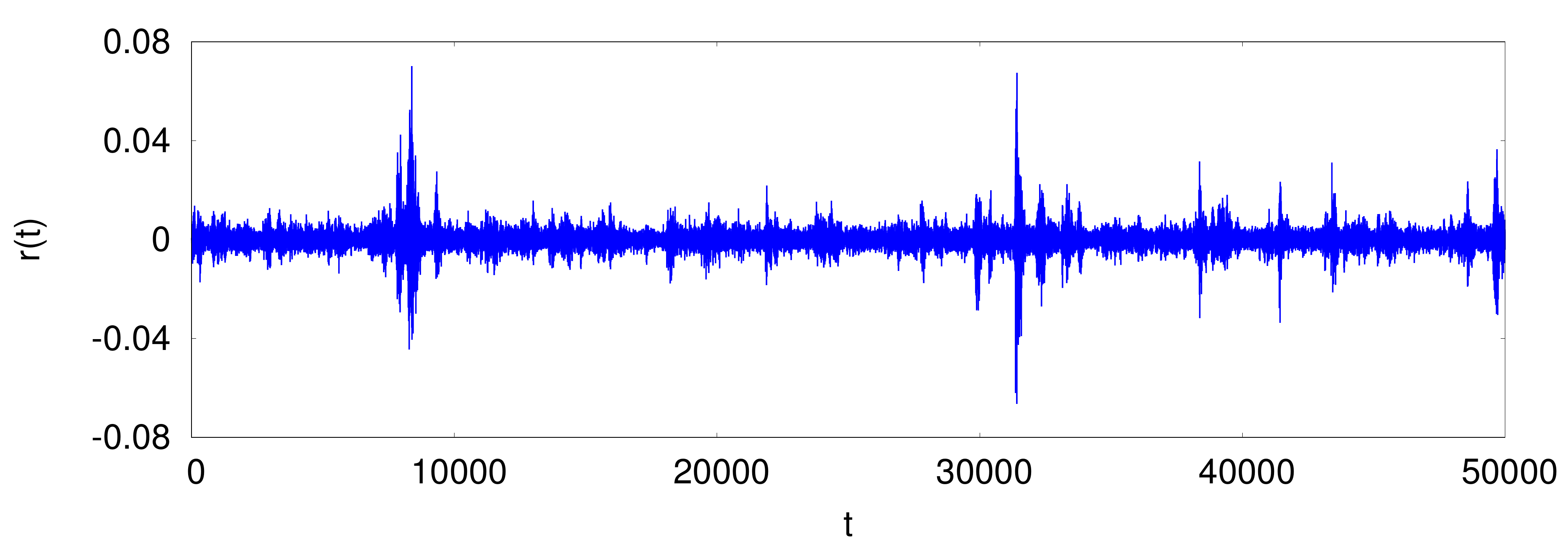}
\end{center}
\caption{The intermittent time series of $r(t)$ displays a temporal structure of volatility clustering in Speculation Game.}
\label{fig2}
\end{figure}

Speculation Game succeeded in reproducing volatility clustering and intermittency. Figure \ref{fig2} displays a yielded time series of $r(t)$, where irregular eruptions with large volatilities clustering together can be observed. Note that there are also some cases (under different parameter settings) in which the system enters the extreme state instead of the volatility clustering state (see Appendix B).

\subsection{Heavy tails}
The distribution of price returns is known to have a heavier tailed character than Gaussian distribution does. The cumulative distribution function  is defined as
\begin{equation}
F_r(x) = {\rm Pr}[r(t) \leq x],
\label{eq12}
\end{equation}
which seems to display a power-law or Pareto-like tail with tail index $\alpha$, which is finite ($2 < \alpha < 5$) for most data sets studied \footnote{Nevertheless, most power-law functions found in previous empirical studies were visually fitted with the least-square method, so it requires to be double-checked by the more reliable statistical techniques recently proposed by Clauset, Shalizi, and Newman \cite{clauset2009power}.} \cite{cont2001empirical}. Note that the probability distribution becomes L\'evy-stable distribution with an infinite variance when $0 < \alpha < 2$ and Gaussian distribution when $\alpha = 2$ \cite{mantegna1999introduction, gopikrishnan2000scaling}.

\begin{figure}[htbp]
\begin{center}
\includegraphics[width=0.7\textwidth]{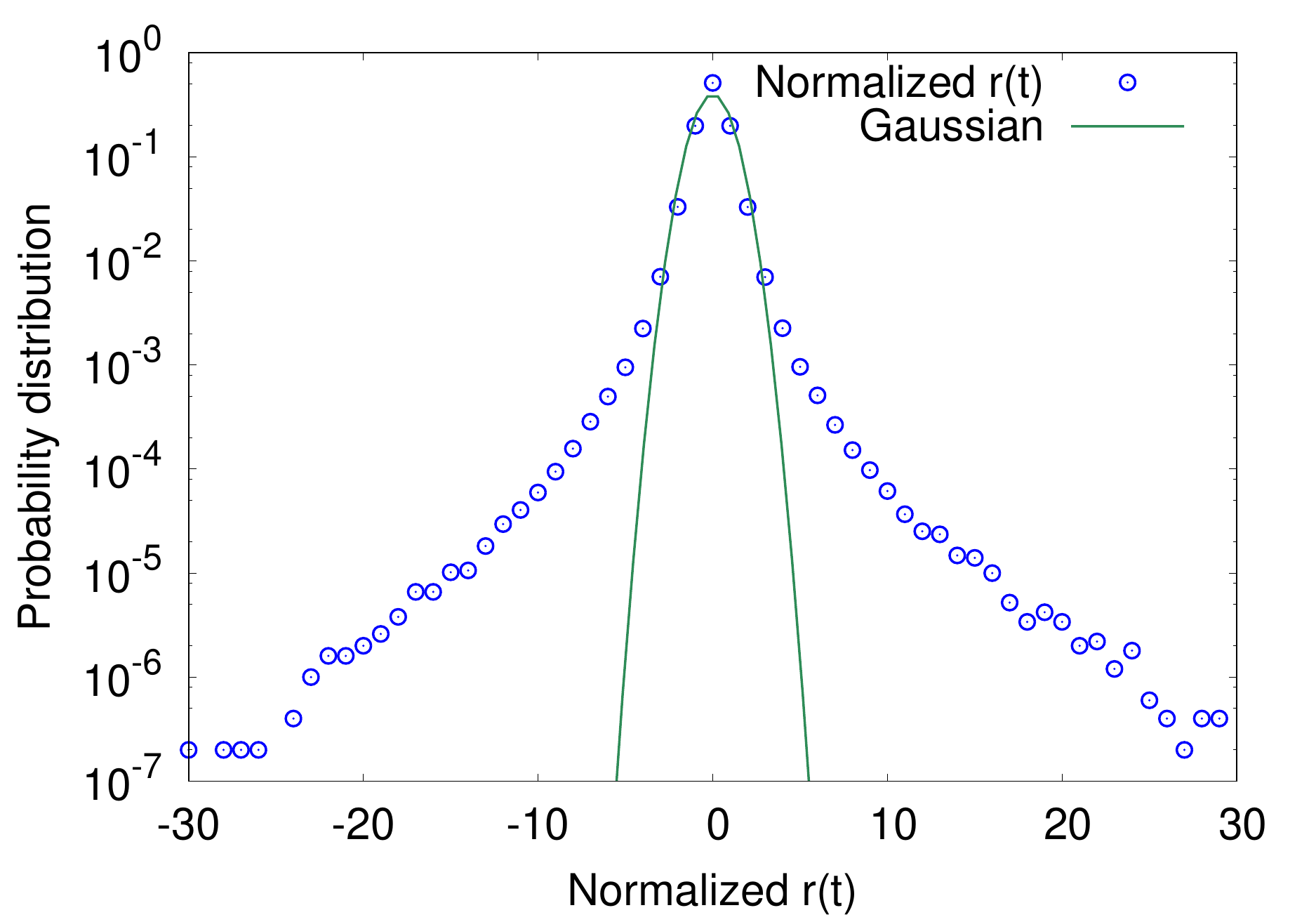}
\end{center}
\caption{The distribution of price returns shows heavier tails than those of Gaussian distribution.}
\label{fig3}
\end{figure}

In Figure \ref{fig3}, the blue circled dots show the averaged probability distribution of market price returns for 50,000 time steps obtained from 100 trials of the simulation of Speculation Game. The returns are normalized to plot Figure \ref{fig3}, in which the vertical axis is in log scale. The return distribution obtained in Speculation Game evidently possesses a fatter tail than that of Gaussian distribution, which is shown as the green line in the same figure.

\begin{figure}[htbp]
\begin{center}
\includegraphics[width=0.7\textwidth]{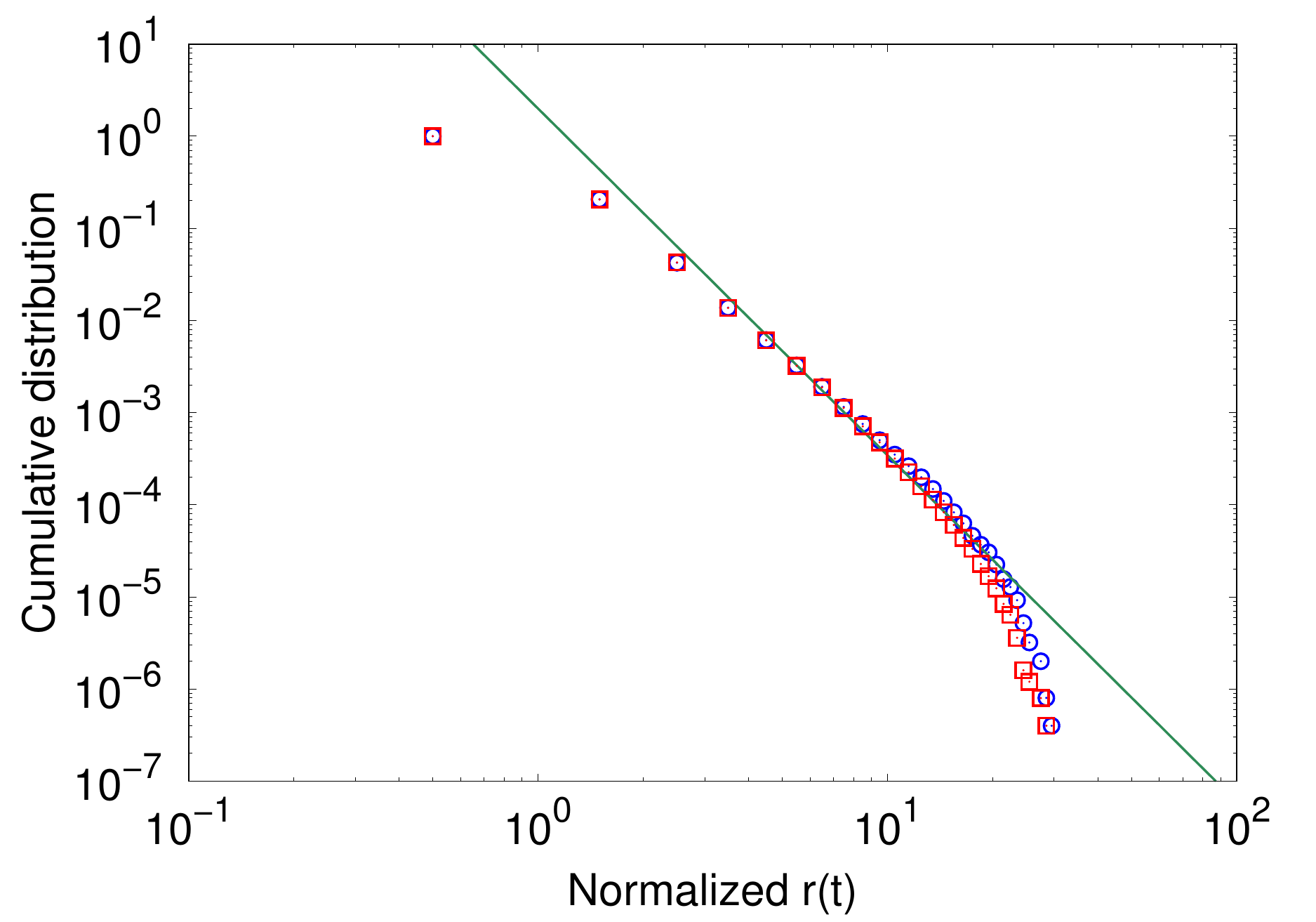}
\end{center}
\caption{Cumulative distributions of price returns displays an asymptotic power law. The blue circles stand for the cumulative distribution of the positive tail and the red squares for the negative tail.}
\label{fig4}
\end{figure}

Moreover, the averaged (complementary) cumulative distributions of its positive and negative tails, $F_r^+(x) = {\rm Pr}[r(t) \geq x]$ and $F_r^-(x) = {\rm Pr}[r(t) \leq x]$, shown as the blue circles and red squares in Figure \ref{fig4}, seems to be subject to a power law. The green line is the asymptotic power-law function with $\alpha \simeq 3.8$, which stays in the range of tail exponents: $2 < \alpha < 5$, and agrees with the inverse cubic law of returns \cite{gopikrishnan1998inverse}.

\begin{figure}[htbp]
\begin{center}
\includegraphics[width=0.7\textwidth]{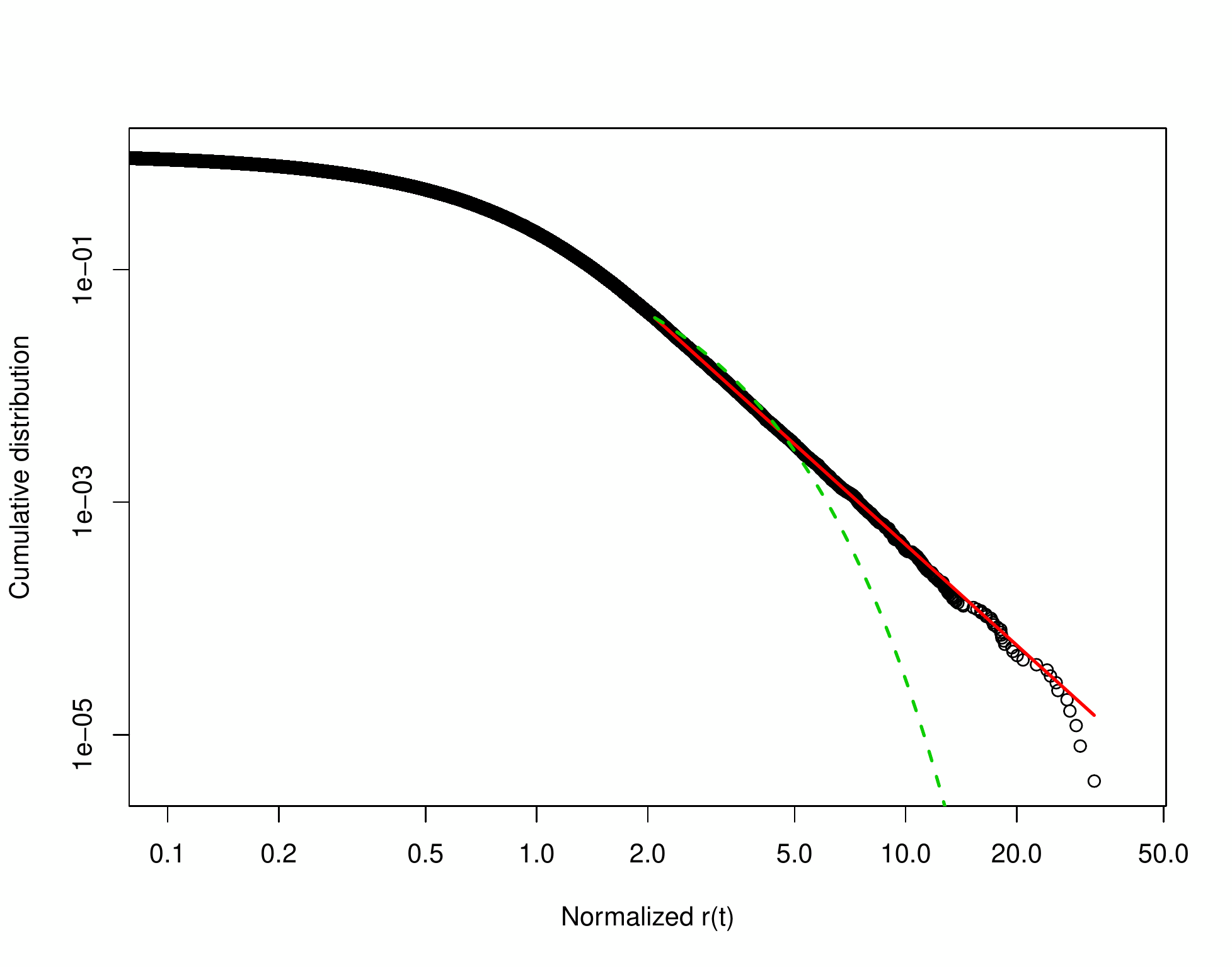}
\end{center}
\caption{The power-law function (red line) fits the cumulative distribution of the positive returns better than the exponential function (green dashed line) does. These results are obtained from the ``poweRlaw'' package \cite{Gillespie2015} analysis of 249,327 data generated by 10 runs of simulations. }
\label{fig5}
\end{figure}

To check the robustness of power-law distribution, we also conducted Vuong's test \cite{vuong1989likelihood} to compare it with the exponential distribution as a competing alternative, by using the ``poweRlaw'' package \cite{Gillespie2015}. This package contains R functions for fitting, comparing and visualizing heavy-tailed distributions, which are based on the statistical techniques proposed by Clauset et al \cite{clauset2009power}. Since its calculation time expands exponentially as the data size increases, we only analyze the normalized positive returns generated from 10-trial simulations in this study.  Figure \ref{fig5} is drawn by finding the lower cut-off of asymptotic power-law function with the package.  The red line is a fitted power-law function with $\alpha = 3.869$, whereas the green dashed line is a fitted exponential function. The red line apparently fits better than the green dashed line to the distribution. The likelihood ratio is $LR = 10.70$ and the p-value is $p=0.00$ ($p<0.1$). Hence, Vuong's test also concludes that the power-law distribution is closer to the true one. 

\subsection{The absence of autocorrelation in returns}
Asset returns in liquid markets typically do not exhibit any significant autocorrelation, defined as
\begin{equation}
\rho_r(\tau) = {\rm Corr}(r(t+\tau), r(t)),
\label{eq13}
\end{equation}
where $\tau$ is the time lag. $\rho_r(\tau)$ exponentially decays within very small time scales ($\simeq$ 20 min) \cite{cont2001empirical, mantegna1999introduction}. The absence of significant linear autocorrelation in returns is often cited as a support for the {\it efficient market hypothesis} \cite{malkiel1970efficient}, as arbitrage opportunities are consumed rapidly.

\begin{figure}[htbp]
\begin{center}
\includegraphics[width=0.7\textwidth]{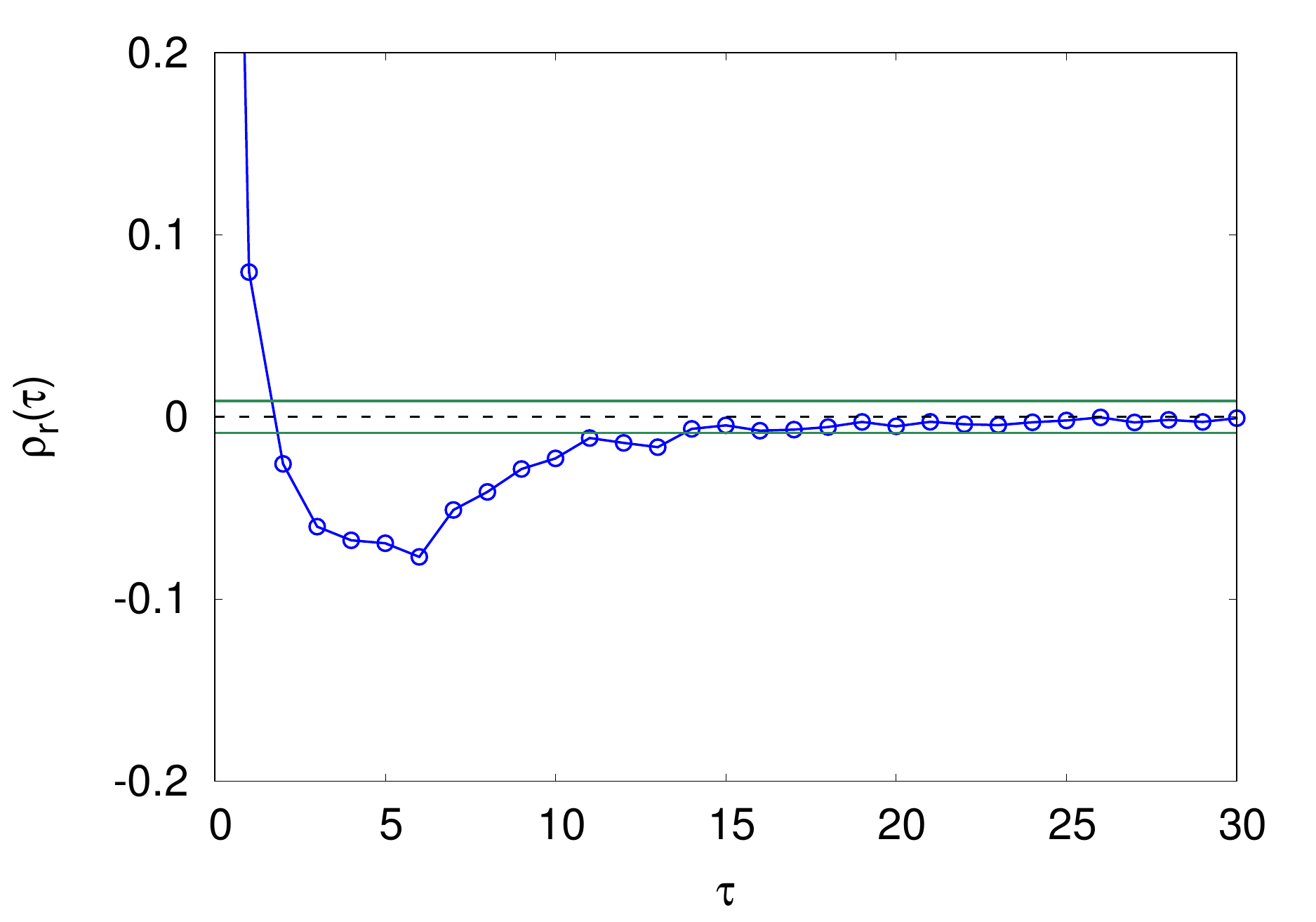}
\end{center}
\caption{The autocorrelation of price returns decays more rapidly than the autocorrelation of volatilities. It enters into the noise zone (between the green lines) at 14 time lags. This result is the average of 100 trials of the simulation for 50,000 time steps.}
\label{fig6}
\end{figure}

Figure \ref{fig6} demonstrates that market price returns yielded by Speculation Game do not display substantial autocorrelation. The two green lines represent the 95\% confidence interval of a Gaussian random walk. The correlation decays within this confidence interval from $\tau > 13$. 

\subsection{Slow decay of autocorrelation in volatilities}
In contrast to fast decay of autocorrelation of returns, the autocorrelation function of absolute returns (volatilities) decays slowly as a function of the time lag $\tau$, defined by
\begin{equation}
\rho_v(\tau) = {\rm Corr}(|r(t+\tau)|, |r(t)|).
\label{eq14}
\end{equation}
The preceding empirical studies show that the decay of $\rho_v(\tau)$ seems to roughly obey either a power law \footnote{This power-law behavior also needs to be reexamined by the recent statistical techniques of Clauset et al \cite{clauset2009power}.} \cite{cont2001empirical, mantegna1999introduction} or logarithmic one \cite{zumbach2007riskmetrics}; there is no consensus yet in the literature about the functional form of the decaying $\rho_v(\tau)$ \cite{zumbach2013characterizing}. The slow decay can be interpreted as a sign of long-range dependence (long memory), causing volatilities to cluster together according to their magnitudes.

\begin{figure}[htbp]
\begin{center}
\includegraphics[width=0.7\textwidth]{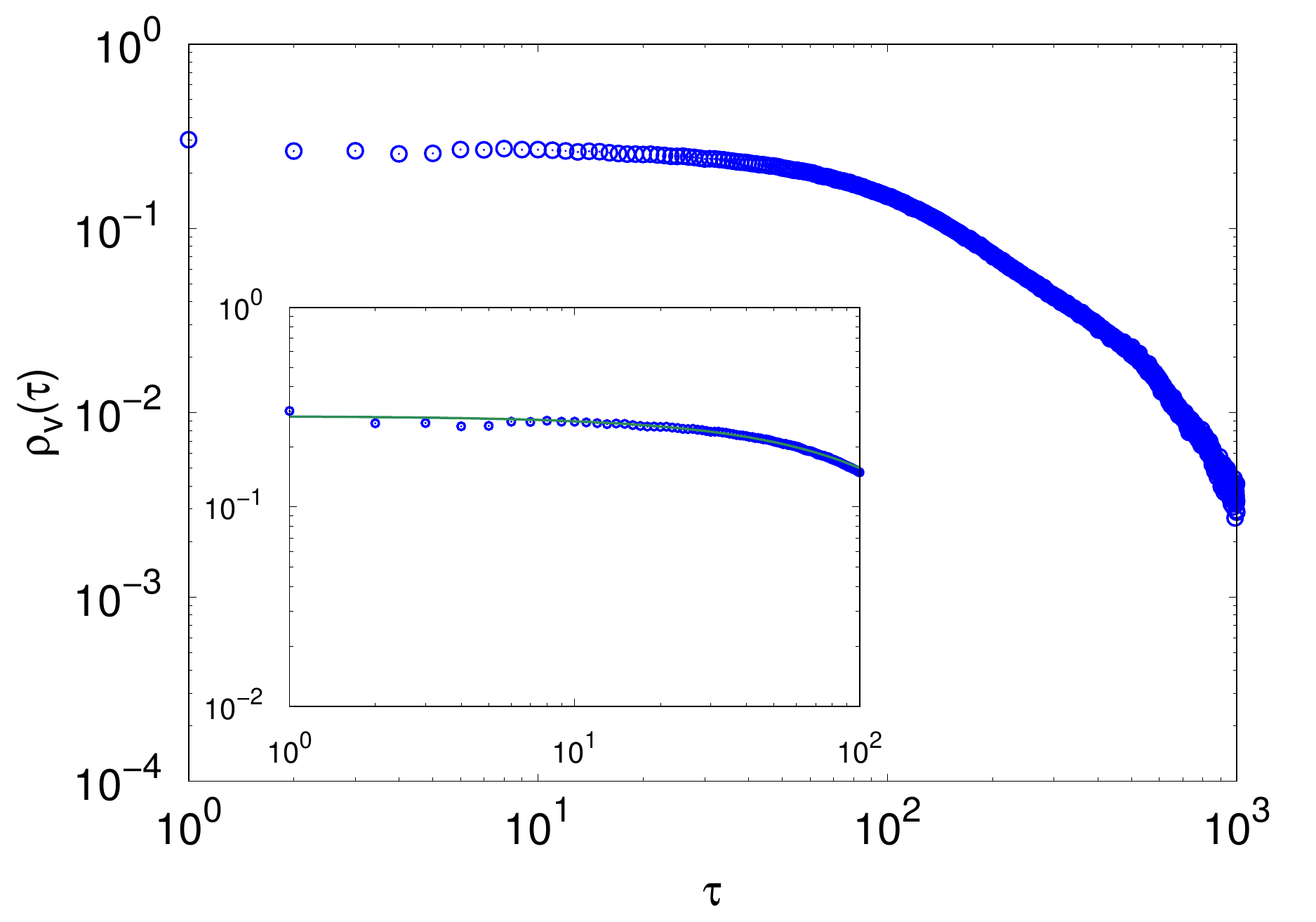}
\end{center}
\caption{The slow decay of autocorrelation of volatilities. Inset: The first 100-time lags decay can be regressed to an exponential function: $\rho_v(\tau) = 0.2853\mathrm{e}^{-0.006\tau}$, as the green line shown.}
\label{fig7}
\end{figure}

Figure \ref{fig7} displays the 100-trial averaged autocorrelation of absolute returns with the same data sets used for Figure \ref{fig6}. The figure indicates that correlations remain positive till $\tau = 10^3$, however, $\rho_v(\tau)$ decays exponentially with an exponent: $6.0\times 10^{-3}$ for $\tau \leq 100$, which is comparable to the statistical analysis of real data. For example, the decay exponents of autocorrelation of absolute return is $5.0\times 10^{-3}$ for gold (daily index: 1978.12.29 -- 2018.8.10) and $7.0\times 10^{-3}$ for IBM stock (daily price: 1962.1.2 -- 2018.8.17).

\subsection{Volume/volatility correlation}
The trading volume is well known to correlate positively with all measures of volatility \cite{bessembinder1993price, chen2001dynamic}, implying that larger price movements are accompanied by higher trading volumes. 

By allowing orders with variable quantities, Speculation Game enables us to express the correlation between the averaged trading volume and volatility for an arbitrary time horizon. The total quantities of buy and sell orders at time step $t$, $Q_{buy}(t)$ and $Q_{sell}(t)$, are calculated respectively as
\begin{equation}
Q_{buy}(t) = \sum_{i=1}^{N}q_i(t)\delta  _{a_i(t), 1},
\qquad
Q_{sell}(t) = \sum_{i=1}^{N}q_i(t)\delta  _{a_i(t), -1},
\label{eq15}
\end{equation}
where the generalized Kronecker deltas are defined as follows:
\begin{equation}
\delta  _{a_i(t), 1} = \begin{cases}
0 & (a_i(t) \neq 1)\\
1 & (a_i(t) = 1)
\end{cases},
\qquad
\delta  _{a_i(t), -1} = \begin{cases}
0 & (a_i(t) \neq -1)\\
1 & (a_i(t) = -1)
\end{cases}.
\label{eq16}
\end{equation}
The averaged trading volume $\langle V(t)\rangle$ for time scale $\Delta t$ can be calculated as follows:
\begin{equation}
\langle V(t)\rangle =max\{ \langle Q_{buy}(t)\rangle, \langle Q_{sell}(t)\rangle \},
\label{eq17}
\end{equation}
where $\langle \rangle$ stands for averaging over the given time scale. 

\begin{figure}[htbp]
\begin{center}
\includegraphics[width=0.7\textwidth]{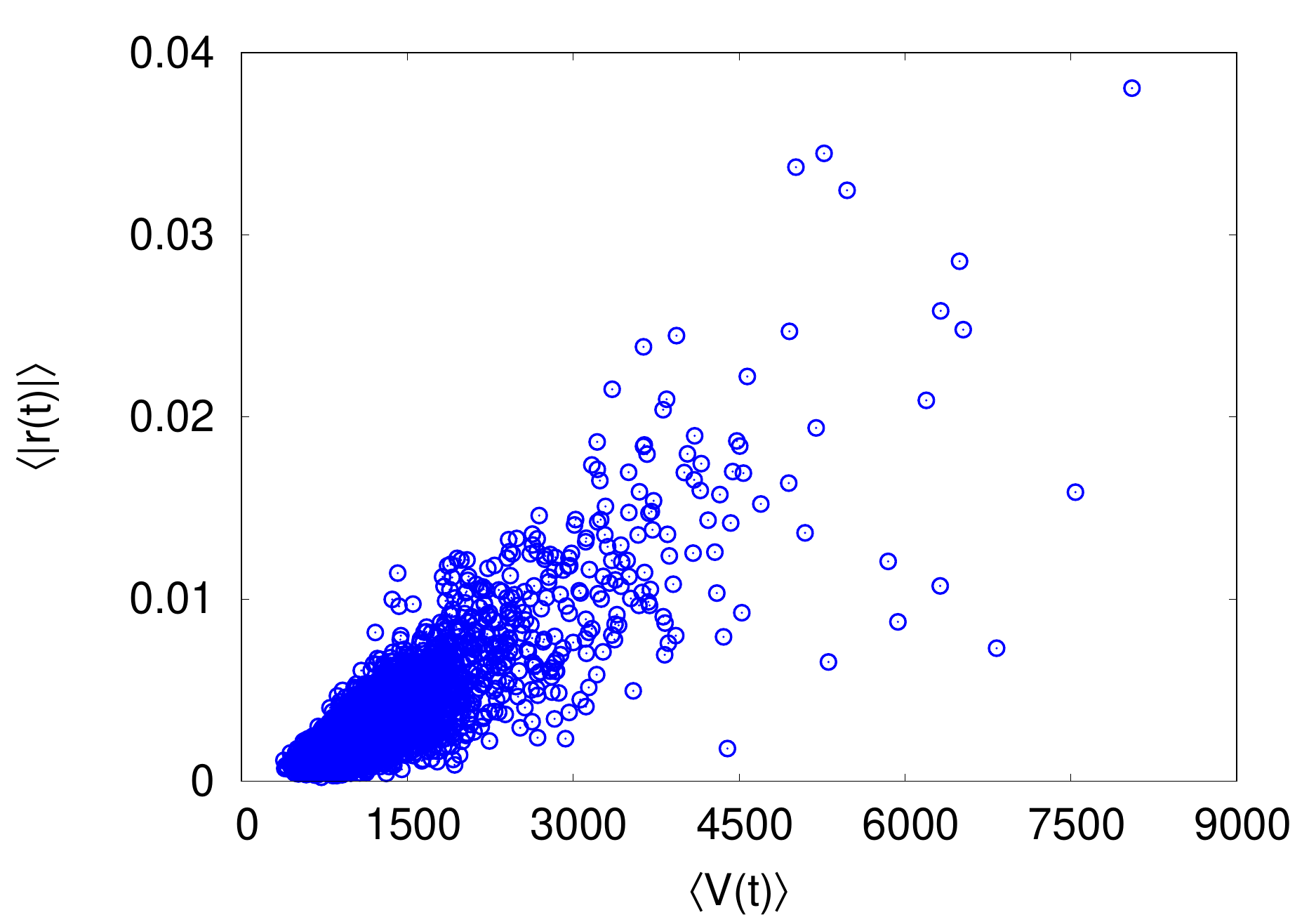}
\end{center}
\caption{The positive correlation between the averaged trading volumes and volatilities in Speculation Game.}
\label{fig8}
\end{figure}

Figure \ref{fig8} is a scatter plot between the averaged trading volume and the averaged volatility for $\Delta t=5$ with the same data sets as used for Figure \ref{fig2}. The figure displays their strong correlation with a coefficient: ${\rm Corr}(\langle V(t)\rangle, \langle|r(t)|\rangle) \simeq 0.84$, showing the capability of Speculation Game in reproducing one of the major stylized facts. Note that the correlation tends to become stronger as $\Delta t$ increases. 

\subsection{Aggregational Gaussianity}
The distribution of returns approaches Gaussian distribution as the time scale $\Delta t$, over which the returns are calculated, increases \cite{rydberg2000realistic, antypas2013aggregational, amien2015inference}. In other words, the shape of the distribution is not identical at different time scales.

The excess degree of kurtosis can be used to describe quantitatively the difference in forms of distribution in reference with Gaussian. Assuming that the return calculated at time scale $\Delta t$ is $r(t,\Delta t)$, its kurtosis $\kappa(\Delta t)$ is defined as follows using its standard deviation $\sigma (\Delta t)$:
\begin{equation}
\kappa(\Delta t) = \frac{\langle (r(t,\Delta t) - \langle r(t,\Delta t) \rangle)^4 \rangle}{\sigma (\Delta t)^4}-3.
\label{eq18}
\end{equation}
The greater positive value of $\kappa(\Delta t)$ means that the return distribution has a sharper peak and fatter tails. Hence, the aggregational Gaussianity can be interpreted as a phenomenon in which the kurtosis $\kappa(\Delta t)$ decreases as the time scale $\Delta t$ increases.

\begin{figure}[htbp]
\begin{center}
\includegraphics[width=0.7\textwidth]{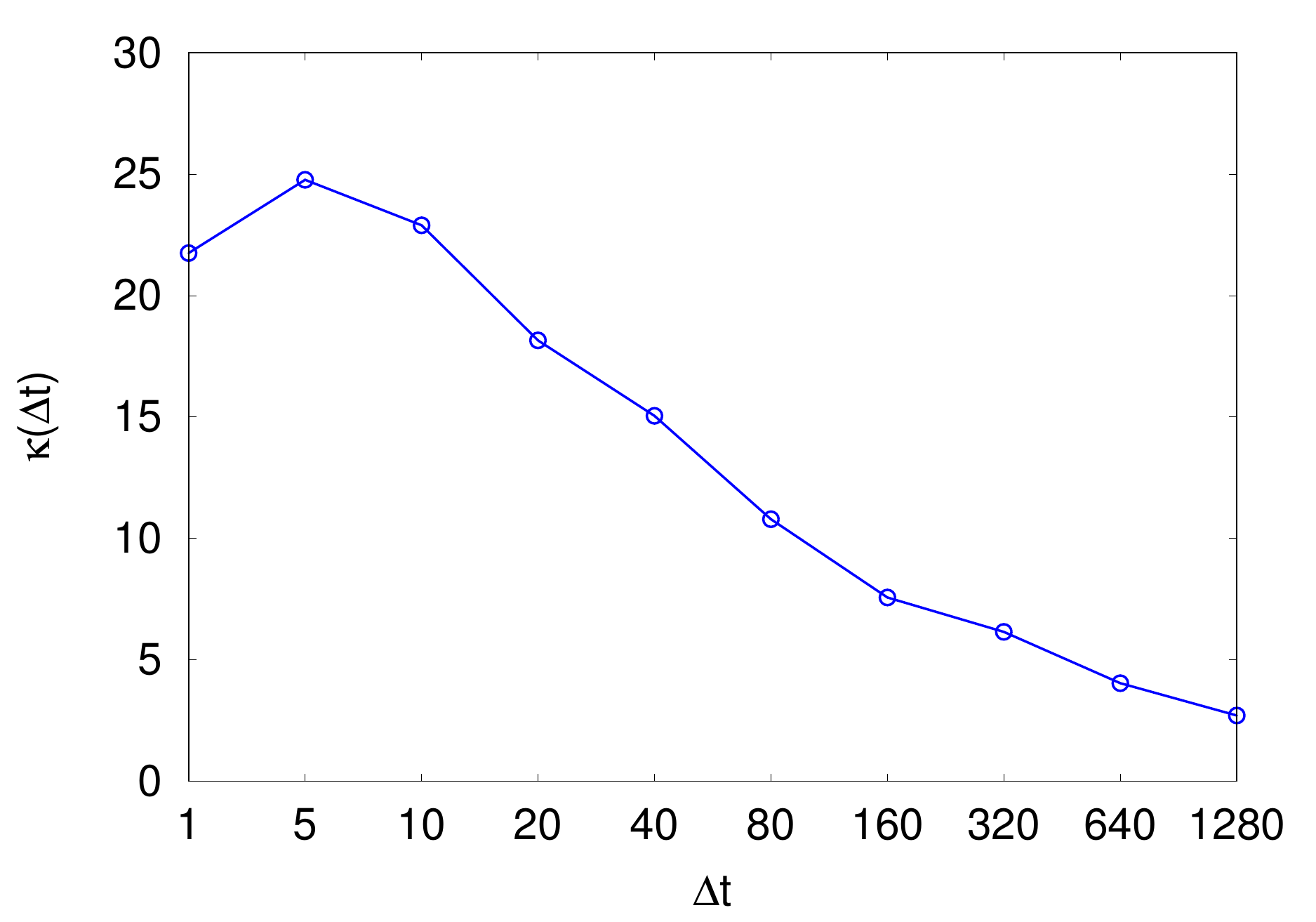}
\end{center}
\caption{The aggregational Gaussianity is recovered in Speculation Game. Each set of data at $\Delta t=5, 10, \cdots, 1280$ is calculated based on the market price returns at $\Delta t=1$ for 50,000 time steps. }
\label{fig9}
\end{figure}

Figure \ref{fig9} shows the change of averaged kurtoses (over 100 trials) by applying different time scales for calculating the returns. In Figure \ref{fig9}, $\kappa(\Delta t)$ tends to decrease as the time scale increases, which seems to be consistent with the observation that the distribution of returns has a power-law tail outside of L\'evy-stable regime. 

Although the result indicates the capability of Speculation Game in reproducing the aggregational Gaussianity, this property is not robust when $\Delta t \leq 20$ in which $\kappa(\Delta t)$ could increase even a 100-trial average is performed. Note that the loss of the aggregational Gaussianity in short time scales is not only a phenomenon appearing in this model, but also an observation in the real market \cite{bos2012spot}. In fact, Johnson et al. \cite{johnson2003financial} had pointed out that a time scale of $\Delta t \geq 60$ min is generally required in order to approximate the probability distribution of return as Gaussian distribution. This is because the assumptions of i.i.d. (independent and identically distributed) return and finite variance do not hold in short time scales. 

\subsection{Conditional heavy tails}
The volatility clustering phenomena may also be recovered to a certain extent by GARCH-type models. However, even after correcting returns for volatility clustering, the probability distribution of the residual time series still possesses heavy tails, which are defined as conditional heavy tails and are less heavy than those in the original distribution of returns \cite{cont2001empirical}.

In a GARCH$(p,q)$ process, return $r_t$ is described with volatility $\sigma_t^2$ as follows \cite{bollerslev1986generalized},
\begin{equation}
\begin{cases}
r_t=\sigma_t\epsilon_t\\
\displaystyle \sigma_t^2=a_0+\sum_{i=1}^{q}a_ir_{t-i}^2+\sum_{j=1}^{p}b_j\sigma_{t-j}^2
\end{cases},
\label{eq19}
\end{equation}
where $p$ and $q$ are non-negative integers, $a_0>0$, $a_i\geq0$ ($i=1, \cdots, q$), $b_j\geq0$ ($j=1, \cdots, p$), and $\epsilon_t \sim \text{i.i.d.} N(0,1)$. To reduce conditional heavy tails, various types of heavy-tailed distributions, such as the Student's {\it t} distribution \cite{bollerslev1987conditionally}, the generalized error distribution (GED) \cite{nelson1991conditional}, and so on are introduced into the GARCH framework \cite{oden2017new}.

\begin{figure}[htbp]
\begin{center}
\includegraphics[width=0.7\textwidth]{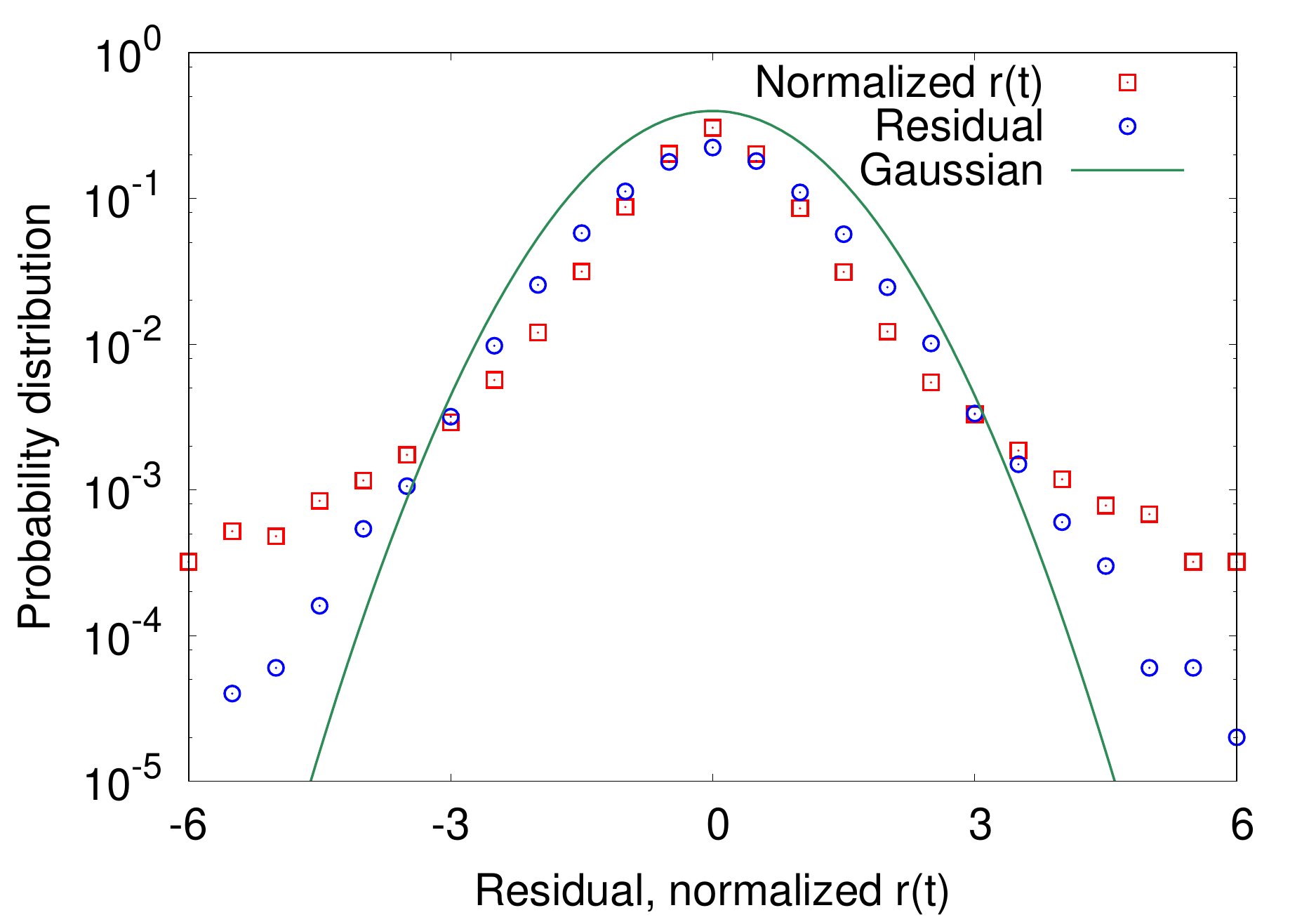}
\end{center}
\caption{Conditional heavy tails observed from the distribution of price returns generated in Speculation Game.}
\label{fig10}
\end{figure}

The price returns from Speculation Game have the conditional heavy tails. Figure \ref{fig10} shows the distribution of normalized returns with the same data sets used for Figure \ref{fig2} (the red squares) and the distribution of its residuals with the GARCH$(1,1)$ model (the blue circles),  against Gaussian (the green line). The residual distribution exhibits fatter tails than those of Gaussian distribution, but thinner than those of the return distribution.

\subsection{Asymmetry in time scales} 
Asymmetry in time scales is a feature that relates to volatilities of different time resolutions. In particular, coarse-grained volatility predicts fine-scale volatility better than the other way around, which indicates that price information flows from the coarse to the fine time scale.

The asymmetry in time scales is described by the lead--lag correlation with respect to the time lag $\tau$ between the two time resolutions. First, coarse volatility $v^c(t)$ and fine volatility $v^f(t)$ are defined as follows with time scale $\Delta t$ \cite{rydberg2000realistic, muller1997volatilities},
\begin{equation}
v^c(t_i) = \left|\sum_{j=1}^n r(t_{i-1}+j\Delta t, \Delta t)\right|,
\label{eq20}
\end{equation}
\begin{equation}
v^f(t_i) = \frac{1}{n} \sum_{j=1}^n \left|r(t_{i-1}+j\Delta t, \Delta t)\right|.
\label{eq21}
\end{equation}
$\Delta t$ has a relationship as $\Delta t= (t_i-t_{i-1})/n$ and $v^f(t)$ can also be defined without $1/n$ to take an average \cite{gavrishchaka2003volatility}. For example, when $\Delta t=1$ d and $n=5$, $v^c(t)$ and $v^f(t)$ represent the weekly based volatility and the (averaged) daily based volatility, respectively. 

The lead--lag correlation $\rho_{cf}(\tau)$ between $v^c(t)$ and $v^f(t)$ is calculated as follows:
\begin{equation}
\rho_{cf}(\tau) = {\rm Corr}(v^c(t+\tau), v^f(t)).
\label{eq22}
\end{equation}
If one plots $\rho_{cf}(\tau)$ against $\tau$ with market data, the function of $\rho_{cf}(\tau)$ will be linearly asymmetric with respect to the $\tau = 0$ line. To show the asymmetry quantitatively, the difference of correlations $\rho_{cf}(\tau)-\rho_{cf}(-\tau)$ and the 95\% confidence interval of a Gaussian random walk are compared with each other. In particular, the negative difference of correlations is out of the confidence interval when $|\tau|$ is small ($|\tau|=1$, or 2). In other words, the correlation between past coarse and future fine volatilities ($\rho_{cf}(\tau)$ when $\tau<0$) is meaningfully bigger than that between past fine and future coarse volatilities ($\rho_{cf}(\tau)$ when $\tau>0$). Accordingly, the relationship of the lead--lag correlation indicates that coarse volatilities can predict fine ones more effectively. 

\begin{figure}[htbp]
\begin{center}
\includegraphics[width=0.7\textwidth]{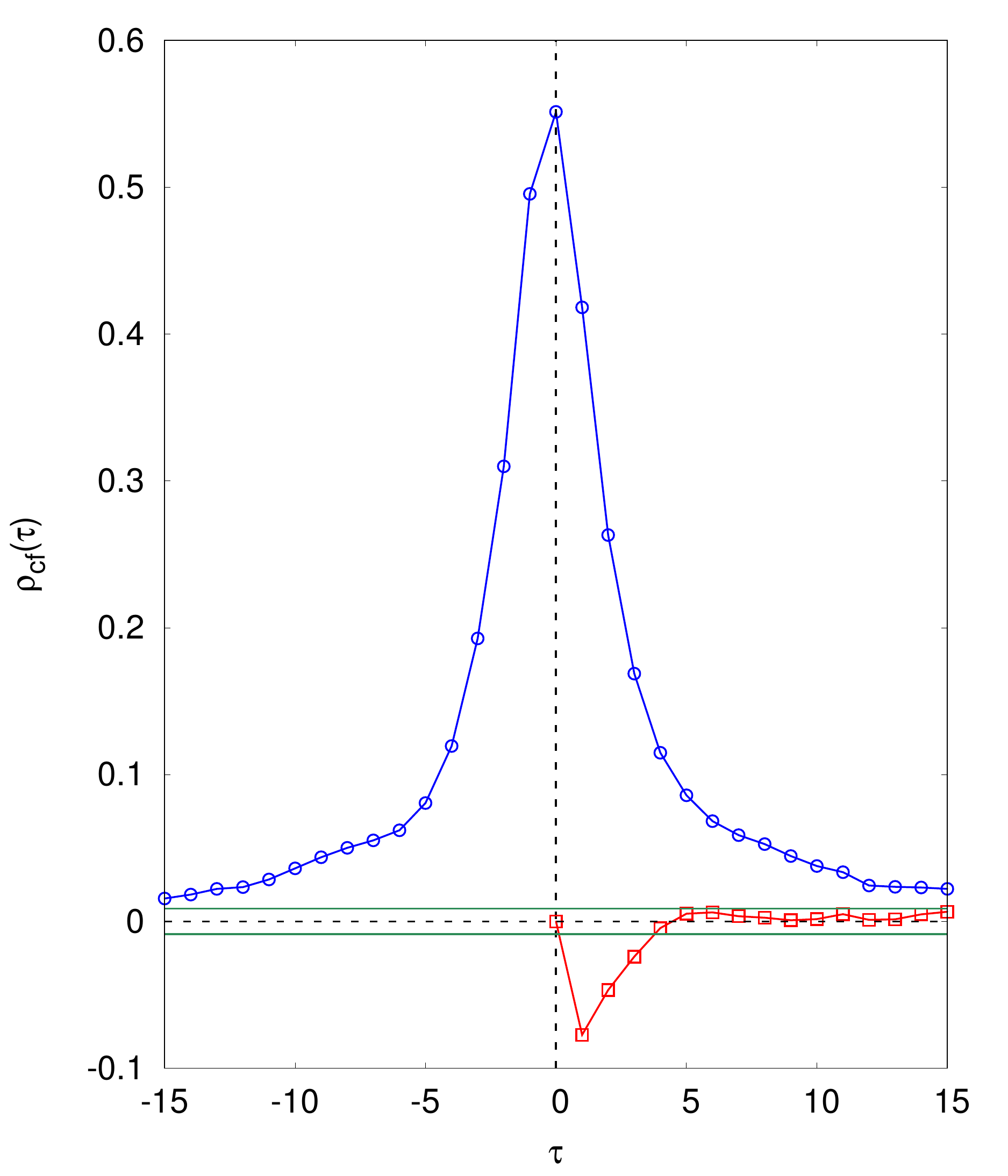}
\end{center}
\caption{Asymmetric lead--lag correlation of coarse and fine volatilities in Speculation Game. Blue circles represent the lead-lag correlation, and red squares the difference of correlations. The 95\% confidence interval of a Gaussian random walk is indicated as the zone between two green lines.}
\label{fig11}
\end{figure}

The asymmetry in time scales can also be observed in Speculation Game. Figure \ref{fig11} displays $\rho_{cf}(\tau)$ with $v^c(t)$ and $v^f(t)$ calculated at regular intervals of 50 time steps by setting $\Delta t=10$ and $n=5$. The lead-lag correlation is calculated with the use of the market price returns for 50,000 time steps and averaged over 100 trials. The linear asymmetry of $\rho_{cf}(\tau)$ (the blue circles) with respect to $\tau = 0$ can be found in Figure \ref{fig11}. In addition, the difference of correlations $\rho_{cf}(\tau)-\rho_{cf}(-\tau)$ (the red squares at the bottom) is outside of the 95\% confidence interval of a Gaussian random walk (the area between the two green lines) when $|\tau|\leq 3$. 

\subsection{Leverage effect}
The leverage effect is a property by which future volatilities are negatively correlated with past returns \cite{bouchaud2001leverage, bouchaud2001more, ahlgren2007frustration}. The leverage effect is measured by the following function with time lag $\tau$,
\begin{equation}
L(\tau) = \frac{\langle [r(t+\tau)]^2 r(t) \rangle}{\langle r(t)^2 \rangle^2}.
\label{eq23}
\end{equation}
$L(\tau)$ is intrinsically the same as ${\rm Corr}(|r(t+\tau)|^2, r(t))$ but has higher measuring sensitivity than ${\rm Corr}(|r(t+\tau)|^2, r(t))$ does, since it is calculated without subtracting the respective averages, and the dimensions of the numerator and denominator are different. In addition, unlike the correlation coefficient, $L(\tau)$ takes a value less than $-1$ and more than $1$.

If $L(\tau)$ is plotted against $\tau$ using real market data, $L(\tau)$ would behave differently crossing the $\tau=0$ boundary. $L(\tau)$ generally takes negligible values when $\tau<0$, though the empirical study for stock indices \cite{bouchaud2001leverage} shows large positive correlations up to $|\tau|=4$ d. On the other hand, when $\tau>0$, $L(\tau)$ takes significant negative values near $\tau=0$ and gradually increases to zero as $\tau$ increases, meaning that there is a relatively large negative correlation between returns and volatilities when the time lag is small. Note that the leverage correlation is stronger but decays faster for stock indices than for individual stocks \cite{bouchaud2001more}.

\begin{figure}[htbp]
\begin{center}
\includegraphics[width=0.7\textwidth]{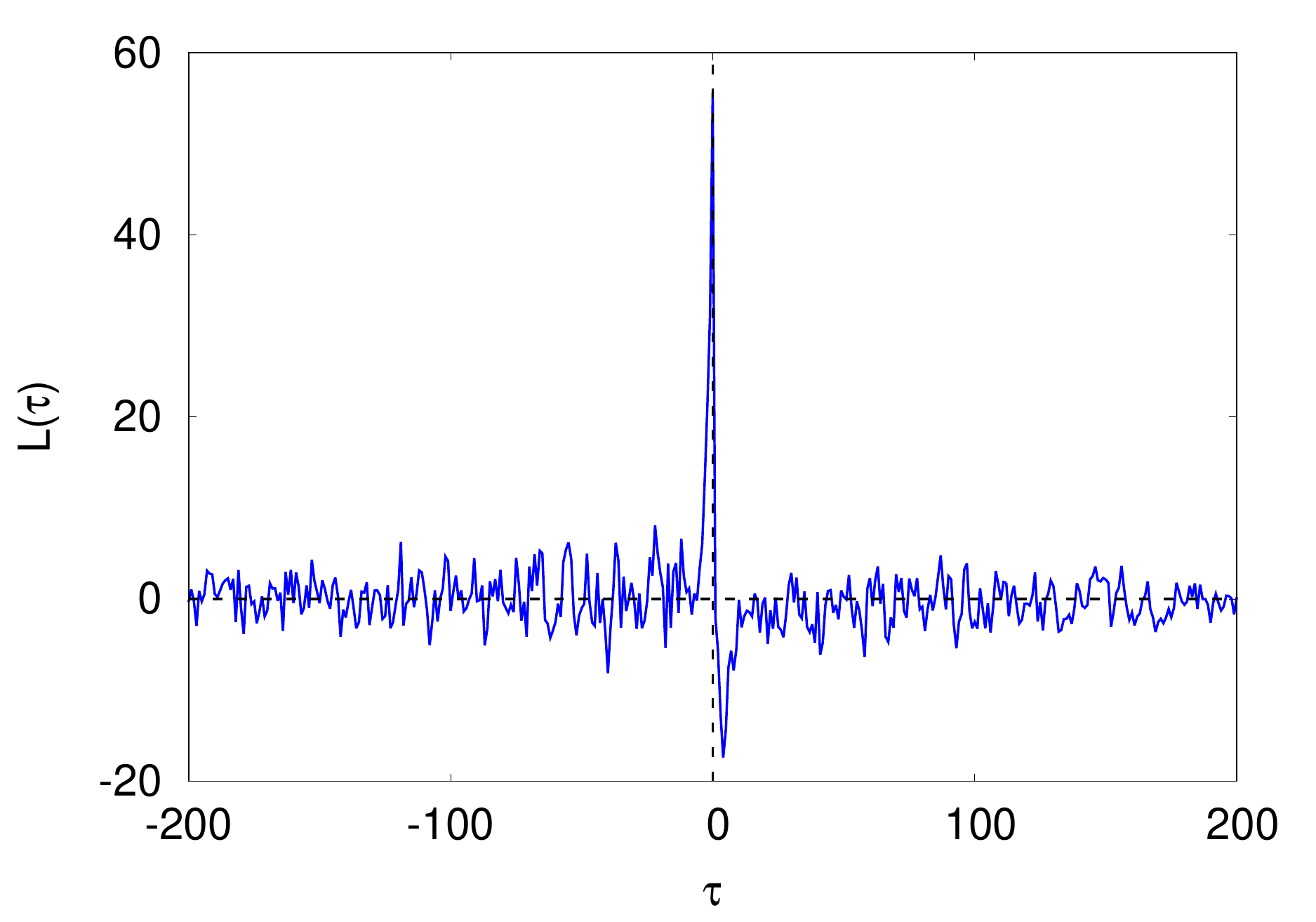}
\end{center}
\caption{The leverage effect reproduced in Speculation Game.}
\label{fig12}
\end{figure}

The leverage effect is successfully reproduced in Speculation Game. Figure \ref{fig12} displays the 100-trial averaged $L(\tau)$ calculated by the market price returns for 50,000 time steps. As shown in the figure, for positive values of $\tau$, $L(\tau)$ drops almost up to $-20$ when $\tau$ is small. Meanwhile, $L(\tau)$ tends to be scattered around zero for negative values of $\tau$. Interestingly, we also find some significant positive correlations up to about $|\tau|=5$, which is quite similar to the result of empirical study \cite{bouchaud2001leverage}. 

\subsection{Gain/loss asymmetry}
Gain/loss asymmetry refers to the difference in the range of movements between upward and downward directions in stock prices and stock index values. Note that this trait is not observed in foreign exchange rates that have a higher symmetry in up/down moves \cite{cont2001empirical}. 

The {\it inverse statistics} method is used to confirm the gain/loss asymmetry \cite{ahlgren2007frustration, jensen2003inverse, donangelo2006synchronization}. This method studies the change in time by the change in price whereas normal financial statistics usually conduct the analysis the other way around. In particular, one decides a return level $\theta$ arbitrarily and measures the least time steps $T_{\pm \theta}(t)$ required for the market price $p(t)$ to reach $p\pm \theta$. With the return $r(t,\Delta t)$, $T_{\pm \theta}(t)$ can be calculated as follows,
\begin{equation}
T_{\pm \theta}(t) =
\begin{cases}
{\rm inf}\{\Delta t \mid r(t,\Delta t) \geq +\theta \}\\
{\rm inf}\{\Delta t \mid r(t,\Delta t) \leq -\theta \}
\end{cases}.
\label{eq24}
\end{equation}
$T_{\pm \theta}(t)$ is called an {\it investment horizon} \cite{simonsen2002optimal}, since it is conceivable as the necessary time to achieve a desirable price change. If one calculates the occurring probability of $T_{\pm \theta}(t)$ for all time steps, an asset featuring the gain/loss asymmetry would exhibit a higher probability in the case of $-\theta$ than in the case of $+\theta$, especially in short time ranges. This means that the price decreases more frequently than it increases in relatively short periods, or more simply, the price falls faster than it rises.

\begin{figure}[htbp]
\begin{center}
\includegraphics[width=0.7\textwidth]{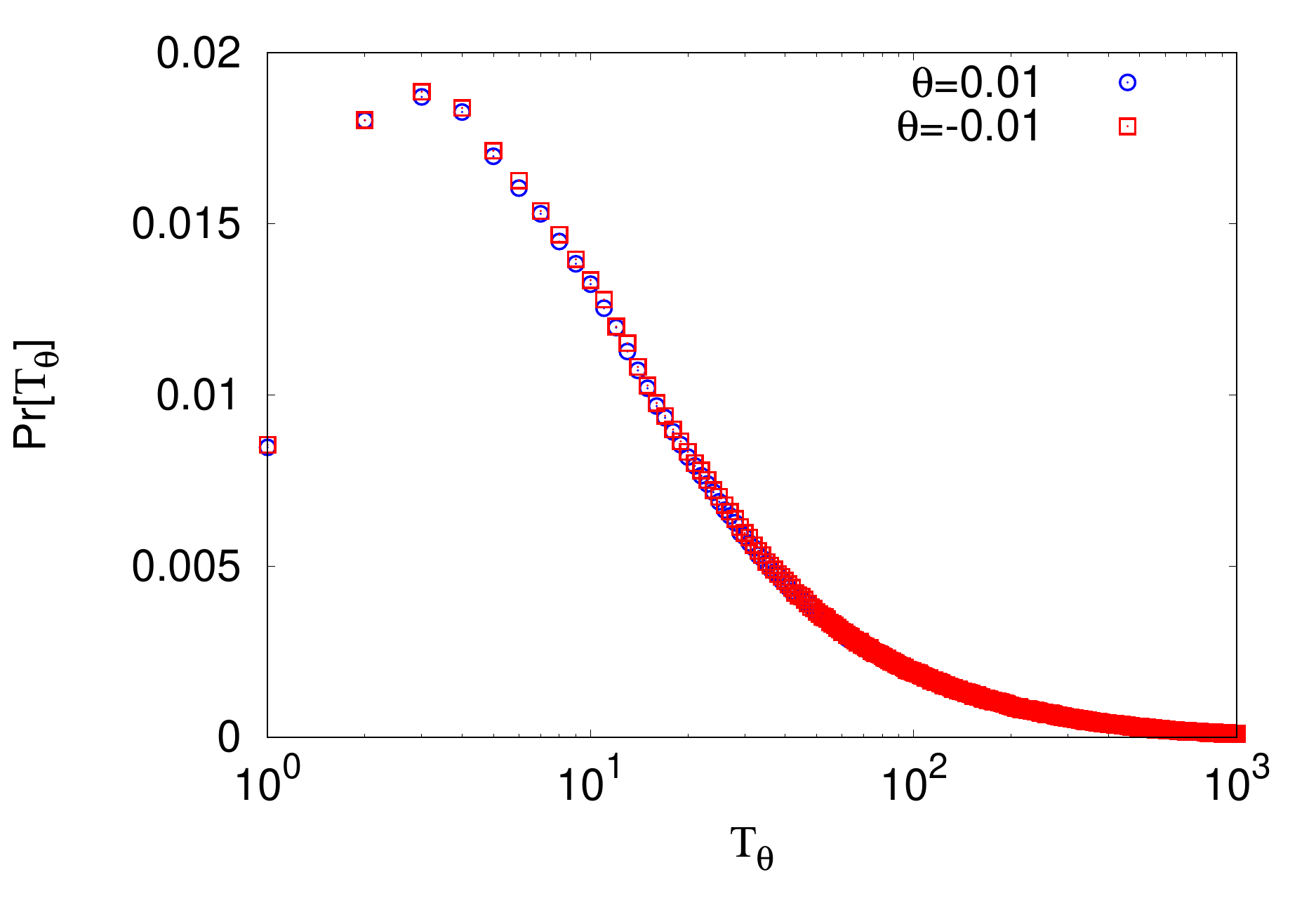}
\end{center}
\caption{Similar distributions of gain/loss investment horizons.}
\label{fig13}
\end{figure}

However, the gain/loss asymmetry cannot be reproduced in Speculation Game. Figure \ref{fig13} shows the 100-trial averaged gain/loss probability distributions ${\rm Pr}[T_\theta]$ for 50,000 time steps by setting the return level as $\theta=\pm0.01$. There is no significant difference in ${\rm Pr}[T_\theta]$ between $\theta=0.01$ (the blue circles) and $\theta=-0.01$ (the red squares), meaning that the gain and loss are symmetric.

The reason for the symmetry of $T_\theta$ distribution probably lies in the symmetric nature of Speculation Game, as the game players are designed to react with no difference either to upward or to downward price movements. In reality, traders may have different reactions to different price movements, in agreement with {\it prospect theory} which states that there is a tendency of risk aversion toward gain or loss of money \cite{kahneman1979prospect}. Therefore, the introduction of such an asymmetric attitude to the model design might be helpful to reproduce the gain/loss asymmetry.

\section{Conclusion}
This study proposes a new, simple agent-based model for the financial markets, which takes account of the speculative behavior pursuing capital gains through round-trip trades. In particular, Speculation Game has three distinguishes itself from other agent based models in three aspects: the enabling of nonuniform holding and idling periods, the inclusion of magnitude information of price change to history, and the implementation of a cognitive world for the evaluation of investing strategies. 

Table \ref{tab2} summarizes the reproducibility of stylized facts for Speculation Game. Symbol ``$+$'' shows that the model successfully recovers the property, and symbol ``$-$'' means that it does not. In the table, it shows that 10 out of 11 stylized facts have been reproduced in the multi-agent simulation with Speculation Game. Note that these reproductions are achieved simultaneously under the same parameter setting. To the best of our knowledge, there exists no other model capable of reproducing the stylized facts at once up to this level, irrespective of the stochastic processes, or the agent-based complex and toy models. Furthermore, we may expect that the remaining unachieved stylized fact could be recovered by introducing the aspect of asymmetric human behavior in the current model.   
\begin{table}[htbp]
\caption{Summary of reproducibility of stylized facts in Speculation Game.}
\label{tab2}
\begin{center}
\begin{tabular}{l|c} \bhline{1.1pt}
Stylized fact & Reproducibility\\ \bhline{1.1pt}
Volatility clustering & $+$ \\
Intermittency & $+$ \\
Heavy tails & $+$ \\
Absence of autocorrelation in returns & $+$ \\
Slow decay of autocorrelation in volatilities & $+$ \\
Volume/volatility correlation & $+$ \\
Aggregational Gaussianity & $+$ \\
Conditional heavy tails & $+$ \\
Asymmetry in time scales & $+$ \\
Leverage effect & $+$ \\
Gain/loss asymmetry & $-$ \\ \bhline{1.1pt}
\end{tabular}  
\end{center}
\end{table}

For future research, it is essential to investigate why and how the stylized properties of asset returns are reproduced in the current model. Although various parameter sets, including the baseline case showed in this study, bring about the state of volatility clustering, with certain parameter combinations (e.g., an overly large memory $M$), Speculation Game would fail to yield any intermittent volatilities, or even enter the extreme state in some cases. Therefore, the parameter conditions for the emergence of stylized facts should be clarified in a more detailed manner. Moreover, the time series displayed in Figure \ref{fig2} reminds one the important property of multifractality. It should be worthwhile to study Hurst exponents of returns at various time horizons or to do more robust multifractal analyses by employing the multifractal detrending moving average (MFDMA) \cite{gu2010detrending} or multifractal detrended fluctuation analysis (MFDFA) \cite{ihlen2012introduction}.

\section*{Acknowledgments}
This work was supported by JSPS KAKENHI grant number JP17J09156. We also would like to thank the reviewers for their valuable comments to improve the quality of our paper and thank Editage (www.editage.jp) for their English language editing services.

\appendix
\setcounter{figure}{0}
\section{The absence of herding behavior}
\begin{figure}[htbp]
\begin{center}
(a)\includegraphics[width=1.0\textwidth]{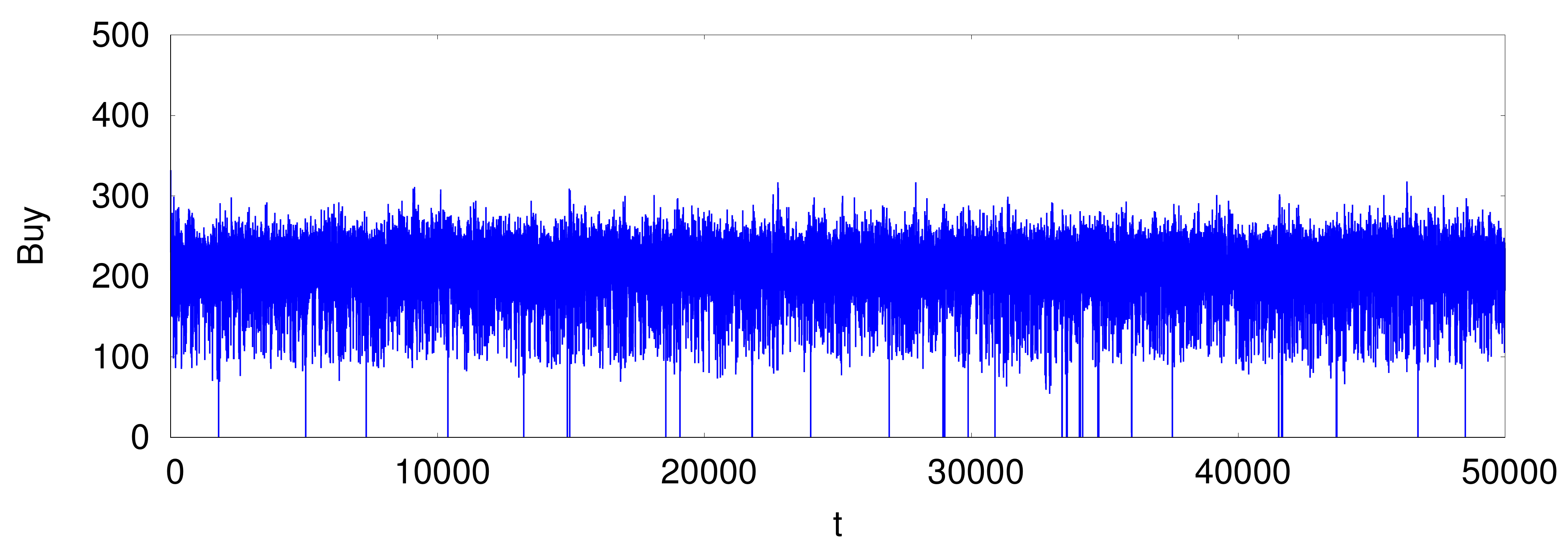}
(b)\includegraphics[width=1.0\textwidth]{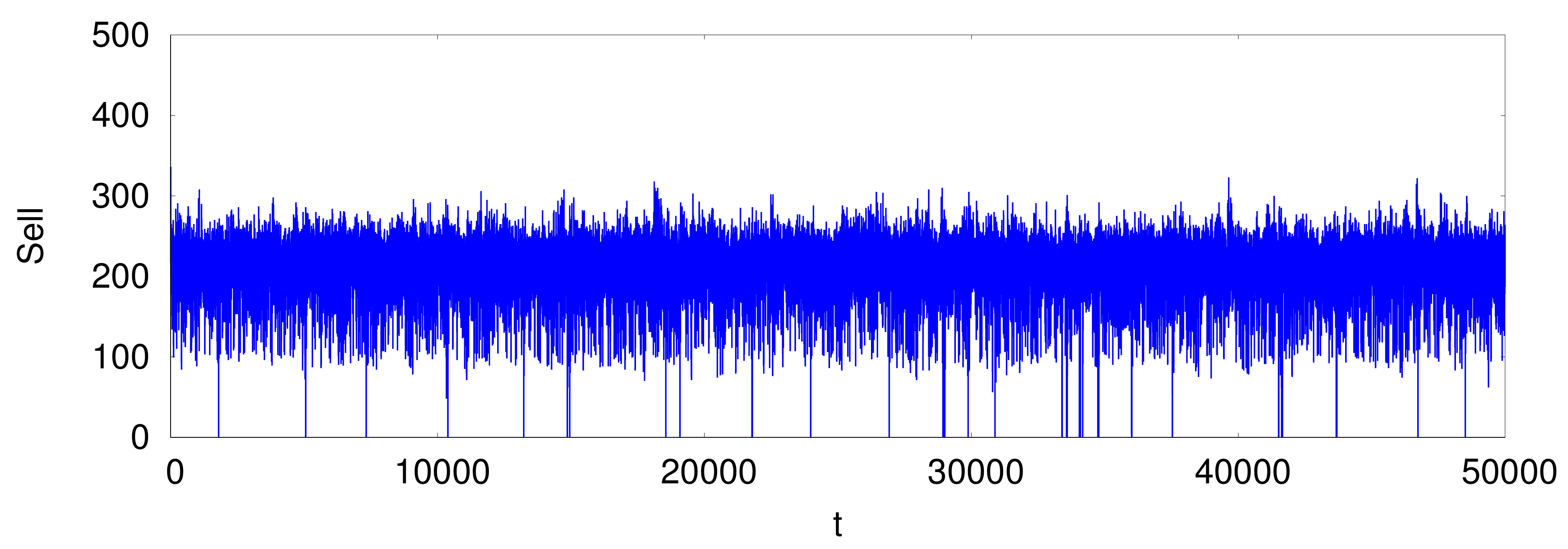}
\end{center}
\caption{The numbers of players who submit (a) buy and (b) sell orders fluctuate constantly for the whole time period. The average numbers are almost constant, which means the number of total ordered players are incessantly present around $N/3 \sim N/2$. }
\label{fig14}
\end{figure}

The absence of herding can be confirmed by observing two time series obtained in the simulation, as shown in Figure \ref{fig14}. These results demonstrate that fluctuations in the numbers of players who submit orders are stable, and the numbers of buyer and seller are almost balanced with each other. The situation does not change much when $M$ varies from 3 to 7. If the herding behavior can be defined as the spontaneously synchronized action of the players in the game, one may say that such a phenomenon is absent in the current model.

\section{The extreme state}
\begin{figure}[htbp]
\begin{center}
\includegraphics[width=1.0\textwidth]{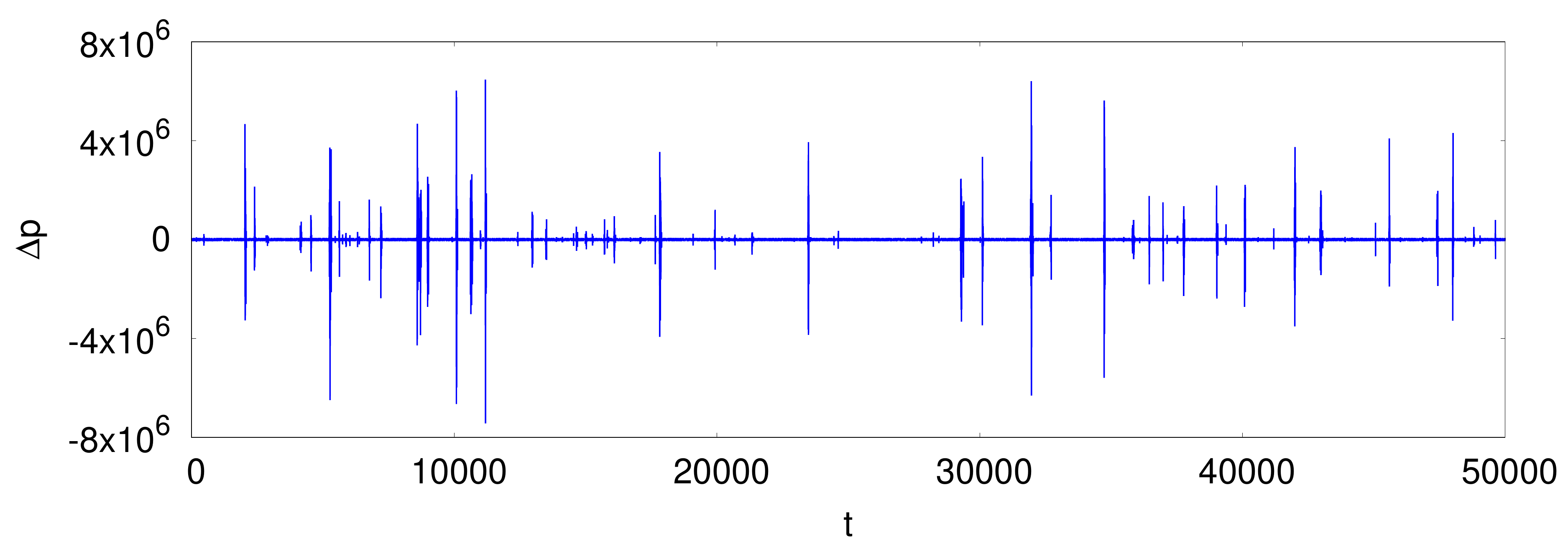}
\end{center}
\caption{Time series of price changes when Speculation Game enters the extreme state ($N=1000$, $M=5$, $S=2$, $B=1$, $C=3$).}
\label{fig15}
\end{figure}

When the board lot amount $B$ is very low, the system will reach the extreme state in which the market price change $\Delta p$ bursts irregularly with tremendous amplitudes, as shown in Figure \ref{fig15}. These tremendous price changes can make the market price $p(t)$ negative, however, the intermittent appearance is almost the same as the state of volatility clustering. 

\end{document}